\documentstyle[11pt,newpasp,twoside,epsf]{article}
\def\edcomment#1{\iffalse\marginpar{\raggedright\sl#1\/}\else\relax\fi}
\reversemarginpar

\begin{document}

\title{{\sc Xmm--newton} AO-1 observations of three SHARC galaxy clusters}
\author{S. Majerowicz, M. Arnaud \& D.M. Neumann}
\affil{$1.$ CEA/Saclay, Service d'Astrophysique, L'Orme des Merisiers, B\^{a}t. 709, 91191 Gif--sur--Yvette Cedex, France}

\begin{abstract}
We present the follow-up of three medium redshift galaxy clusters from the SHARC survey observed with {\sc Xmm--newton}. We studied RX~J0256.5+0006 which shows two components which are very likely in interaction. The smallest component exhibits a comet-like structure indicating ram pressure stripping as it falls onto the main cluster. The second cluster, RX~J2237.0-1516 is an elliptical cluster with a gas temperature of 3.0$\pm$0.5\,keV. The third cluster, RX~J1200.8-0328 seems to be in a relaxed state because its shape is regular and we do not see obvious temperature gradient. Its mean temperature is 5.1$^{+0.7}_{-0.5}$\,keV.
\end{abstract}

\section{Introduction}

Catalogs of galaxy clusters with a large range of redshifts and cluster {\sc x}-ray luminosities are an ideal basis for the test of cosmological parameters (e.g. Henry 2000~; Borgani et al. 2001) and the study of cluster formation and evolution. The Bright and Southern Serendipitous High Redshift Archival {\sc Rosat} Cluster (SHARC) surveys provide a sample of clusters detected in {\sc Rosat} observations over two decades of {\sc x}-ray luminosities (10$^{43}\ <$ L$_{\mathrm{x}}\ <\ 10^{45}$\,erg/s) with redshifts between 0.2 and 0.8 (Romer et al. 2000~; Collins et al. 1997~; Burke et al. 1997).

We present here preliminary results based on the follow-up observations of three of these selected galaxy clusters with {\sc Xmm--newton}~: RX~J0256.5+0006, RX~J2237.0-1516 and RX~J1200.8-0328. For the f\mbox{}irst time, the new generation of {\sc x}-ray observatory, like {\sc Xmm-newton} and {\em Chandra} give capabilities to do precise spectroscopic and imaging analysis at the same time (see e.g Arnaud et al. 2002a).

Our analysis is based on the three {\sc epic} cameras, {\sc mos}1\&2 and pn. Throughout the paper, we use a cosmology with  H$_{0}=50$\,km/s/Mpc, $\Omega_{m}=0.3$ and $\Omega_{\Lambda}=0.7$. The error bars are given with a conf\mbox{}idence level of 90\%.

\section{Data treatment}

All the data presented here are treated with the version 5.2 of the {\sc sas} (Science Analysis Software). We select event patterns and time intervals of low background as described in Majerowicz et al. (2002). Table 1 shows the effective exposure time of our observations after flare rejection.

\begin{table}
\caption{Remaining exposure time in units of ks after f\mbox{}lare rejection}
\begin{center}
\begin{tabular}{cccc}
\hline
Observations & {\sc mos}1 & {\sc mos}2 & pn\\
\hline
RX~J0256.5+0006 & 10.6 & 10.3 & 7.1\\
RX~J2237.0-1516 & 8.7 & 8.7 & 8.7\\
RX~J1200.8-0328 & 28.3 & 28.4 & 22.0\\
\hline
\end{tabular}
\end{center}
\end{table}

We also correct for vignetting of the telescope with the weigthing method fully described in Arnaud et al. (2001).

\section{Background subtraction method}

The background subtraction is necessary for low surface brightness extended sources like galaxy clusters. After removing f\mbox{}lares, the remaining background components are the high energy particle induced background, which is not vignetted by the telescopes, and the cosmic {\sc x}-ray background (hereafter {\sc cxb}) which is vignetted and depends on sky position of the observation (see Snowden et al. 1997).

To remove these components, we apply the technique described in Majerowicz et al. (2002) and Pratt et al. (2001). We use a blank sky event f\mbox{}ile produced for each {\sc epic} instrument (Lumb 2002). We f\mbox{}irst subtract the blank f\mbox{}ield from the source according to a normalisation in the hard energy band. This step allows to subtract the high energy particle induced background. The second step consists in the correction of the {\sc cxb}. This is performed by using data in the region outside the cluster emission. We reproduce the f\mbox{}irst step for this region and the expected residuals are the difference between the local {\sc cxb} and the {\sc cxb} in the blank f\mbox{}ield. We f\mbox{}inally subtract these residuals from the product of the f\mbox{}irst step.

\section{RX~J0256.5+0006}

RX~J0256.5+0006 is a cluster of galaxies at a redshift of 0.36 (1' corresponds to 420\,kpc). F\mbox{}igure 1 indicates a bimodal structure of the cluster. In order to understand the dynamics of the cluster we have to assess whether the bimodal structure is linked to merging or it is due to chance alignment.

This cluster hosts a main component and a smaller second component in the west. Since in a cluster-subcluster merger, the subcluster encounters more impact on its gas distribution, we concentrate on the western component which is clearly the smallest structure. We f\mbox{}it the {\sc x}-ray emission of the main component with an elliptical $\beta$-model (see Neumann \& B\"{o}hringer 1997). This f\mbox{}it is done by excluding point sources and the western cluster. We obtain $\beta=0.82$, a major axis of 370\,kpc and a minor over major axis ratio of 0.80. We subtract this cluster model from the count rate image. The residuals which are shown as contours in f\mbox{}igure 1, are measured in terms of signif\mbox{}icance. The western structure is really apparent at more than 3\,$\sigma$ and display a comet-like form.

\begin{table}
\begin{center}
\begin{tabular}{cc}
{\hspace*{-1cm}\plotfiddle{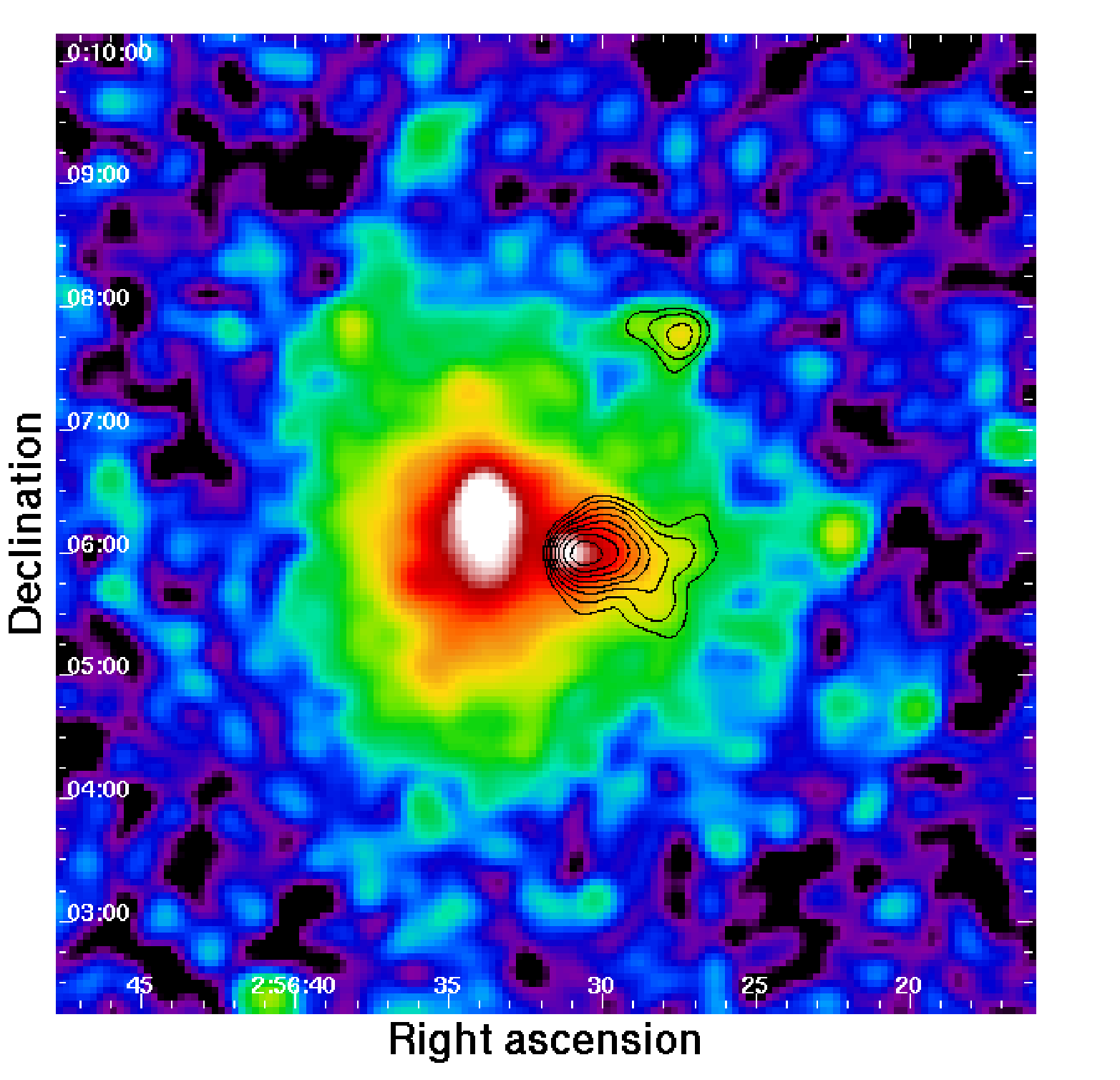}{6cm}{0}{27}{27}{-350}{0}}
&
\plotfiddle{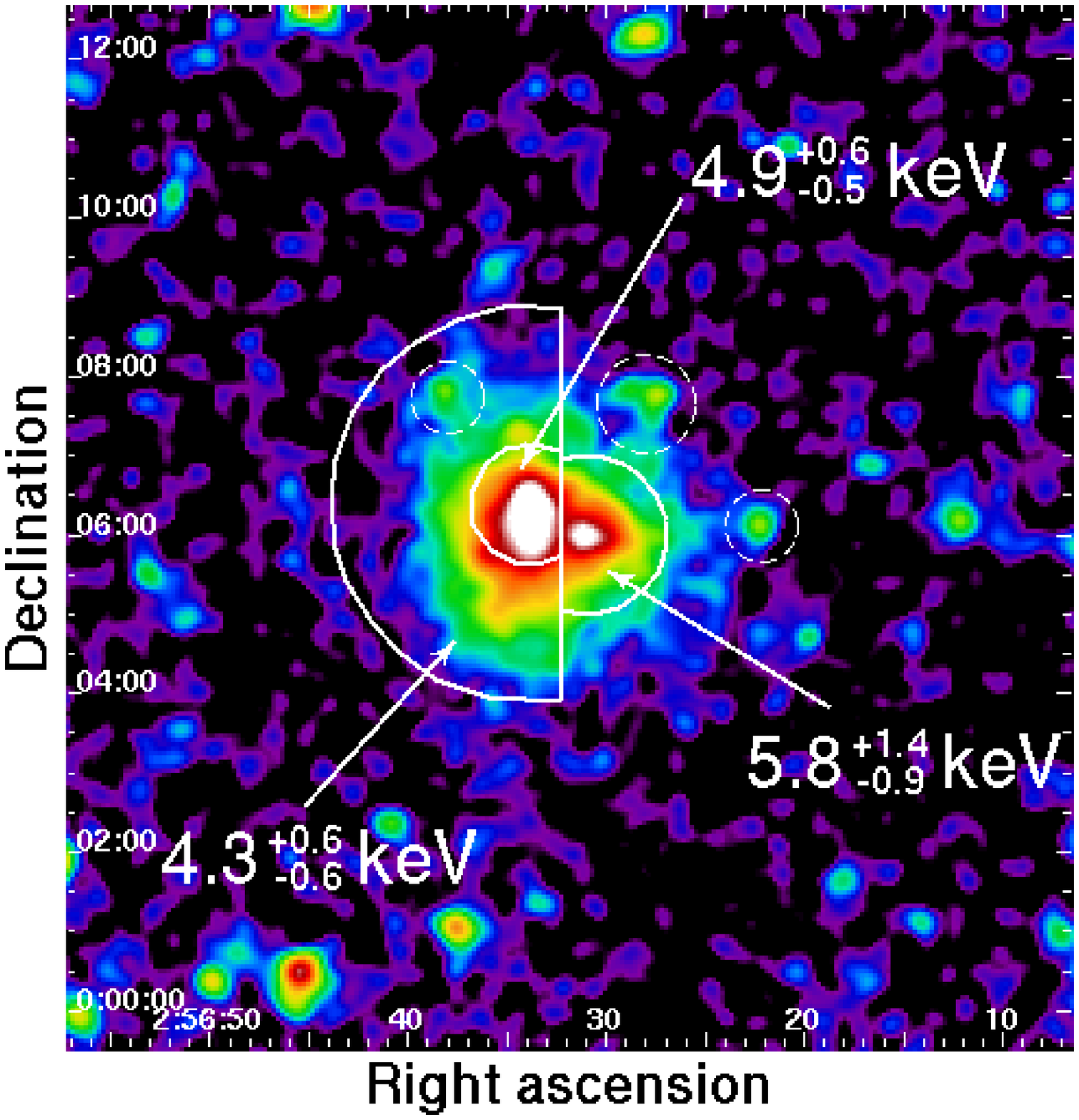}{6cm}{0}{32}{32}{-550}{0}
\end{tabular}
\begin{tabular}{p{14cm}}
Figure 1. {\bf Left~:} Image of RX~J0256.5+0006 in the 0.3 to 3\,keV energy band. The overlayed contours shows the residuals in signif\mbox{}icant $\sigma$s (the lowest contour is 3\,$\sigma$ and stepwidth between two contours is 1\,$\sigma$) after the subtraction of the best f\mbox{}it elliptical $\beta$-model for the main component. {\bf Right~:} Cluster image overlayed with the regions where we measure the temperature (East is on the right and North on the top).
\end{tabular}
\end{center}
\end{table}

The strong compression of the isophots in the direction of the main component and the presence of a tail in the opposite side suggests that these two components are in interaction. It also indicates that the subcluster gas is stripped by ram pressure as it encounters the main cluster atmosphere. Furthermore, simulations made by Roettiger et al. (1997) have shown that gas stripping is more eff\mbox{}icient in case of mergers with large differences in mass of the components which seems to be the case here.

To see possible spatial temperature variations, we measure temperatures in selected regions as shown in f\mbox{}igure 1. All measured temperatures agree relatively well within the error bars.

Moreover, numerical simulations show (Takizawa 1999~; Schindler \& M\"{u}ller 1993) that cluster mergers create major shock fronts or adiabatic compression which heat the intracluster medium between the colliding components. The region between the center of the main component and the comet-like structure has a measured temperature of 5.0$^{+1.6}_{-1.0}$\,keV and suggests that no shock wave has been created. However this region is relatively large (140$\times$700\,kpc) and we cannot thus exclude large temperature gradients at smaller scales.

\section{RX~J2237.0-1516}

RX~J2237.0-1516 is a medium redshift cluster of galaxies (z=0.299). We f\mbox{}irst f\mbox{}it the intra cluster medium distribution with an elliptical $\beta$-model and we f\mbox{}ind $\beta$=0.67, a major axis of 285\,kpc and a minor over major axis ratio of 0.75. After the subtraction of this model, we do not see extended structures with a significance greater than 3\,$\sigma$ in the cluster. This is particularly true for the northern emission extension of the cluster where a point source seems to be detected. Fitting a 1D $\beta$-model to the cluster we obtain $\beta=0.63$ and $r_{c}=$0.55'$=$205\,kpc.

\begin{table}
\begin{center}
\begin{tabular}{c}
\plotfiddle{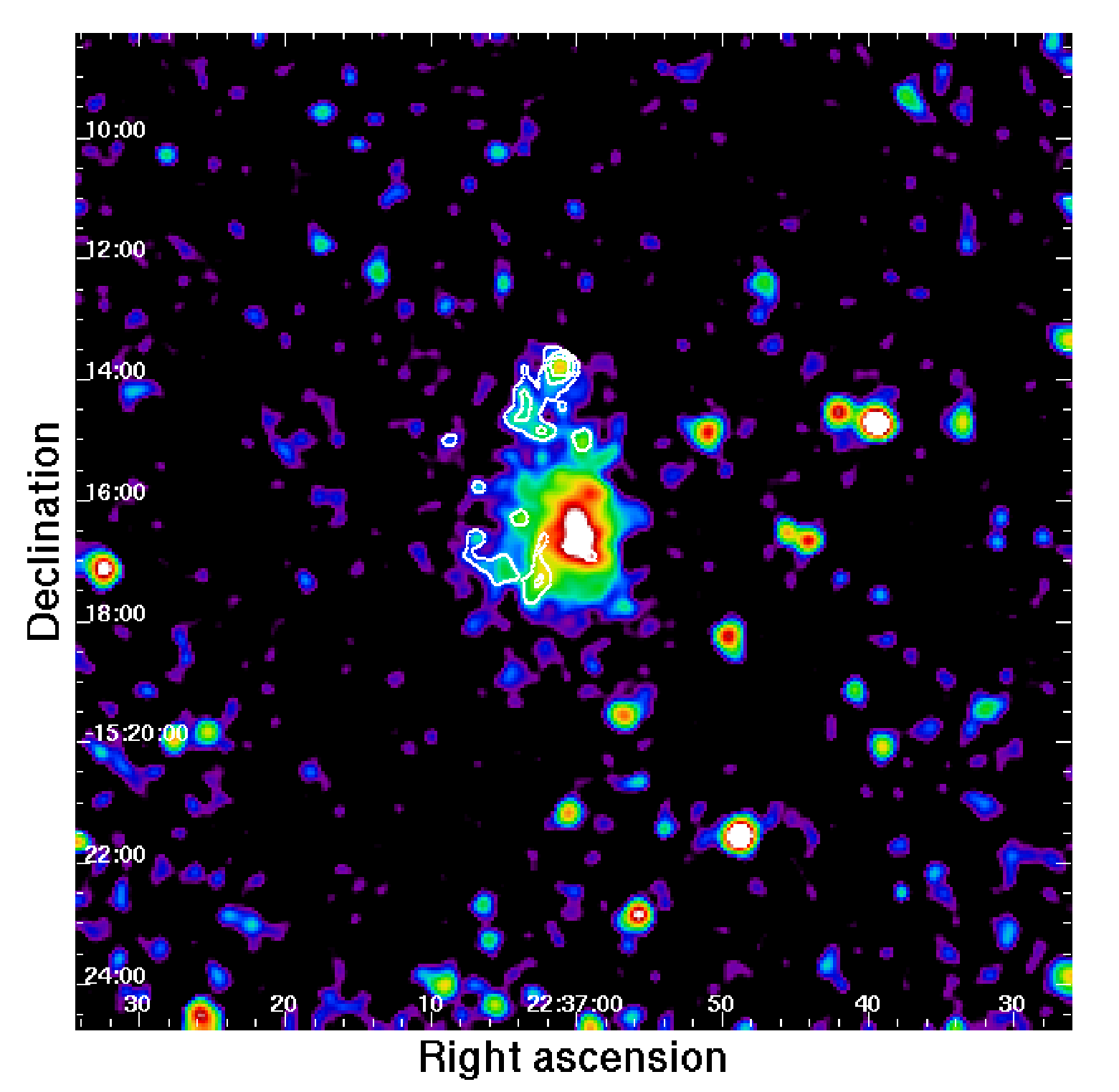}{6cm}{0}{27}{27}{-280}{0}
\end{tabular}
\begin{tabular}{p{10cm}}
Figure 2. Image of RX~J2237.0-1516 in the 0.3 to 3\,keV energy band. The contours are the residuals in signif\mbox{}icant $\sigma$s (the lowest represents 2\,$\sigma$ and the step width is 1\,$\sigma$).
\end{tabular}
\end{center}
\end{table}

Performing spectral f\mbox{}itting, we measure the mean temperature of RX~J2237.0-1516 and we f\mbox{}ind 3.0$\pm$0.5\,keV within a radius of 2' or 750\,kpc centered on the maximum {\sc x}-ray emission peak.

\section{RX~J1200.8-0328}

RX~J1200.8-0328 is at a redshift of 0.395 and with f\mbox{}igure 3, it seems to be spherically symmetric and in a relaxed state. We determine the temperature of the intra cluster medium in two distinct annuli (see f\mbox{}igure 3, the outer radii are 45'' and 2', i.e. 336 and 896\,kpc). The temperature prof\mbox{}ile seems to be fairly flat up to 900\,kpc from the cluster center. 

\begin{table}
\begin{center}
\begin{tabular}{c}
\plotfiddle{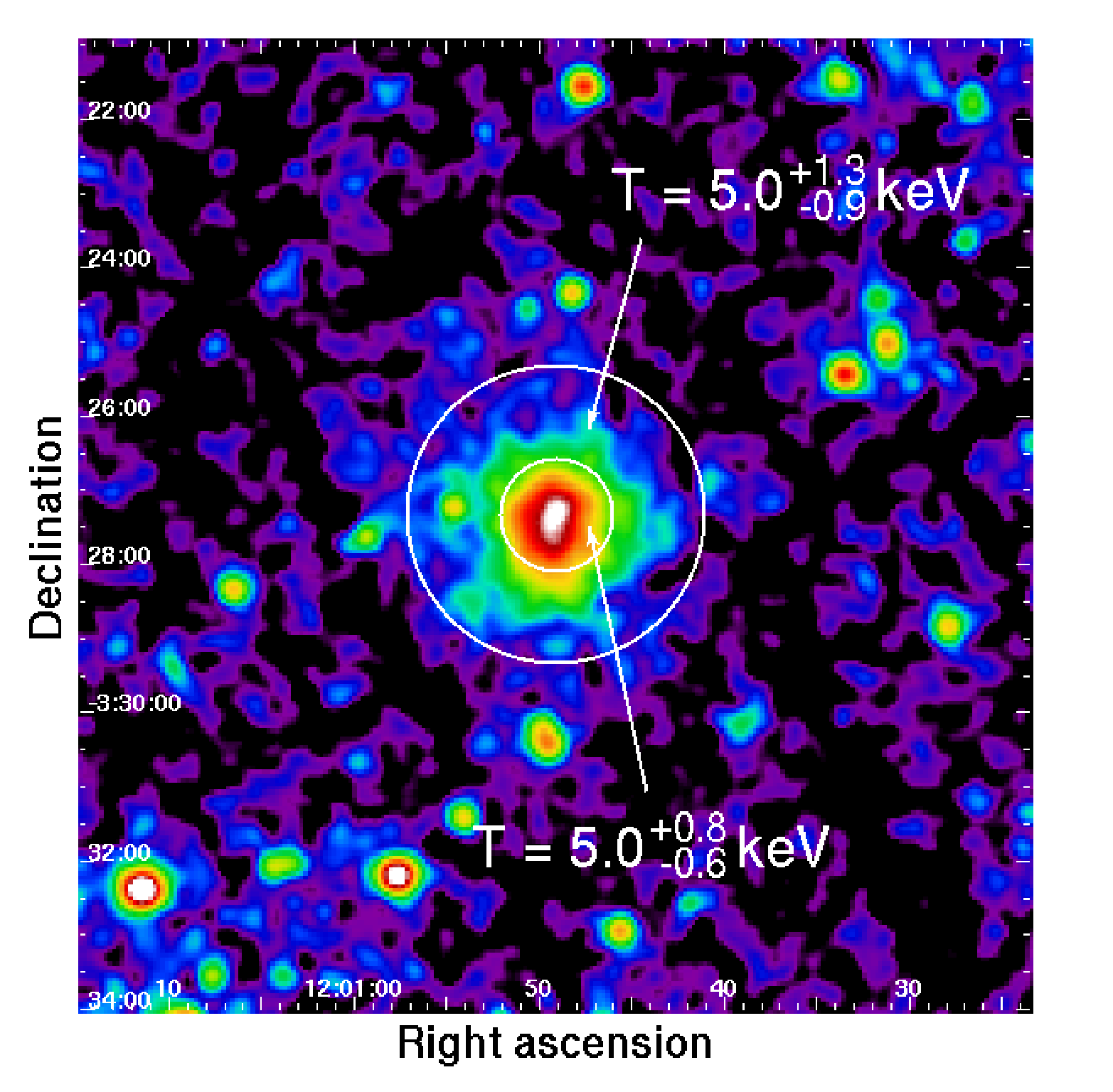}{6cm}{0}{27}{27}{-280}{0}
\end{tabular}
\begin{tabular}{p{10cm}}
Figure 3. Image of RX~J1200.8-0328 in the 0.3 to 3\,keV energy band. The two annuli where we extract temperature estimates are overlayed.
\end{tabular}
\end{center}
\end{table}

This suggests that this cluster does not host a strong cooling flow. This was already seen in RX~J1120.1+4318, a relaxed cluster at z=0.6 (Arnaud et al. 2002a) also found in the SHARC survey. The global temperature for this cluster is 5.1$^{+0.7}_{-0.5}$\,keV. Its surface brightness prof\mbox{}ile is f\mbox{}itted by a $\beta$-model and the parameters are~: $\beta$=0.64 and $r_{c}=$0.51'$=$228\,kpc.

RX~J2237.0-1516 and RX~J1200.8-0328 are also used in the study of their emission measure prof\mbox{}ile derived from their surface brightness profile and their mean temperature (see Arnaud 2002b).

\end{document}